\documentclass[aps,secnumarabic,nobalancelastpage,amsmath,amssymb,
nofootinbib]{revtex4}
\usepackage[cp1251]{inputenc}
\usepackage[T2A]{fontenc}
\usepackage[english]{babel}
\usepackage{amssymb,latexsym,amsmath,amscd}
\usepackage{graphicx,epsfig,amsmath,amsfonts,epstopdf}
\usepackage{physics}
\usepackage{scalerel,bm}
\usepackage{bigints}
\usepackage{mathtools}
\usepackage{subfig}
\usepackage{float}
\usepackage{caption}

\graphicspath{ {images/} }

\DeclareMathOperator*{\rint}{\ThisStyle{\rotatebox{13}{$\SavedStyle\!\int\!$}}}

\begin{document}
\fontsize{11}{11}\selectfont 
\title{Qualitative types of  cosmological evolution in hydrodynamic models with barotropic equation of state}
\author{V.I.~Zhdanov, S.S.~Dylda}
\affiliation  {Taras Shevchenko National University of Kyiv \\  Astronomical observatory,  Observatorna st. 3,\\ 04053  Kyiv, Ukraine}
\email {valeryzhdanov@gmail.com;  tunerzinc@gmail.com}

\begin{abstract}
 We study solutions of the Friedmann equations in case of the homogeneous isotropic Universe filled with a perfect fluid. The main points concern the monotony properties  of the solutions, the possibility to extend the solutions  on all times and occurrence of singularities. We present a qualitative classification of all possible solutions in case of the general smooth barotropic equation of state of the fluid, provided the speed of sound is finite. 
  The list of possible scenarios includes  analogs of the ``Big Rip'' in the future and/or in the past as well as singularity free solutions and oscillating Universes. Extensions of the results to the multicomponent fluids are discussed.
		
	
\end{abstract}
\maketitle
\section{introduction}
\label{introduction}
Modern  $\Lambda CDM$ cosmological model successfully describes most of the
observational data of extragalactic astronomy. Nevertheless, due to the 
well-known horizon and flatness problems, modifications of the standard cosmological model are widely discussed to 
ensure the existence of the inflationary stage of the cosmological evolution. In this view a  dynamical models of the dark energy (DE) have been introduced, which are different from the unchanging cosmological constant. Various DE models involve  cosmological fields, extra dimensions, modified gravity etc (see 
\cite{grt_1}-\cite{dedm} for a review). Often enough to analyze these issues the  hydrodynamic approach is used,
where either all the matter in the Universe or the dynamic DE is modeled by means of a relativistic fluid with some equation of state (EoS). On this way  a number of
analytical solutions have been found (e.g. \cite{Noiri_1}-\cite{big_rip}  and references therein). 

In these studies, a considerable attention is paid to the  qualitative properties of solutions, such as monotonicity, intervals of existence and limiting properties of the solutions. Recently, interest in the solutions like "Big Rip" \cite{grand_rip} and 
in some other types of singular behavior \cite{Noiri_1,worse_big_rip,grand_rip,parnovsky} has grown.  Typical questions are as follows. Does a solution of the Friedmann equations exist for all $t\to \infty$?  Otherwise, does  the cosmological scale factor and/or $e(t)$ blow up at some singularity point? Is the energy density $e(t)$ bounded? 

In paper \cite{jenk_zh}, such a qualitative behavior of solutions has been studied for a special form of  EoS subject to some restrictions. 
In the present paper we relax these restrictions. We consider the homogeneous isotropic Universe with a general barotropic EoS $p = p(e)$, that relates the pressure $p$ to the invariant energy density $e > 0$. The only conditions imposed are the smoothness of the function $p(e)$ and the existence of an upper bound for $dp/de$, i.e. the speed of sound is supposed to be bounded. We describe possible scenarios of the cosmological evolution with a focus on roots of specific enthalpy $h(e) = e + p(e)$. The smoothness of $p(e)$  and $h(e)$ is rather a strict condition; for example, it prohibits crossings of the "phantom line" ($e + p=0$).  We present below a complete list of all possible scenarios with various qualitative behaviors.  

\section{basic equations}

The homogeneous isotropic cosmology is described by the FLRW metric 
\begin{equation}
\label{eq1}
ds^2 = dt^2 - a^2\left( t \right)\left[ {d\chi ^2 + F^2\left( \chi
\right)dO^2} \right] \quad ,
\end{equation}
\noindent
where $F(x) = \sin (x)$, $\sinh (x)$ or $x$ respectively, for the closed, open and spatially-flat Universe. This corresponds to the following values of the parameter $k=1,-1,0$ in the Friedmann equations
\begin{equation}
\label{eq2}
\frac{d^2a}{dt^2} = - \frac{4\pi }{3}a\left( {e + 3p} \right)
\end{equation}
\begin{equation}
\label{eq3}
\left(\frac{1}{a}\frac{da}{dt}\right)^2 = \frac{8\pi }{3}e - \frac{k}{a^2},
\end{equation}

\noindent here $G=c=1$. 
The only non-trivial hydrodynamics equation is 
\begin{equation}
\label{eq4}
\frac{de}{dt} + \frac{3h}{a}\frac{da}{dt} = 0 \quad ;
\end{equation}
Equations (\ref{eq2},\ref{eq3},\ref{eq4}) are not independent, so we use further (\ref{eq3},\ref{eq4}).
Equation (\ref{eq4}) can be rewritten as a first-order autonomous equation
\begin{equation}
\label{eq5}
\frac{de}{dX} = - 3h,
\quad
X = \ln a \quad .
\end{equation}

Further we introduce notations as follows:

$f:a_1 \uparrow a_2 $  means that function $f(x)$ is monotonically increasing from $a_1$ to $a_2>a_1$ when $x$ belongs to the function domain. Analogously $a_1 \downarrow a_2$ in case of the 
decreasing function.

$f:a_1 \uparrow a_2 \downarrow a_1 $ means that the function $f$ is monotonically
increasing from $a_1 $ to $a_2 > a_1 $ and then, after reaching the turning point $a_2 $, it is monotonically decreasing to $a_1 $.

We  denote decreasing unbounded solutions of (\ref{eq5}) by the symbol $\bf {U \downarrow }$, and increasing unbounded solutions by $\bf {U\uparrow }$. Analogously, for increasing  bounded solutions and decreasing bounded ones we write correspondingly  $\bf {B\uparrow }$ and  $\bf {B \downarrow }$.

\section{solutions of the equation (\ref{eq5})}
\label{solutions_of_ref_eq5}

We supposed that $h$ is a smooth function. 
The condition that $dp/de$ is bounded means 
$\exists\, 0<C_0^2<\infty$ such that\footnote{We do not need here $C_0^2\le 1$. }
\begin{equation}
\label{dpde}
|dp/de|\le C_0^2 
\end{equation}
and we have  $|dh/de|\le 1+C_0^2$. The right-hand side of (\ref{eq5}) is Lipschitz continuous and $\forall e\in (-\infty, \infty)$ there exists  the finite Lipschitz constant $3(1+C_0^2)$. Then in virtue of the  Cauchy-Lipschitz  theorem, the equation (\ref{eq5}) with initial data $e(t_0)=e_0$ has a unique smooth solution $e(X)$ for all $X \in (-\infty,\infty)$.

Suppose we have $e_1:h(e_1)=0$, then $e(X)\equiv e_1$ is a solution of equation (\ref{eq5}). In virtue of the uniqueness, any other
solution $e(X)$ of this equation cannot intersect the line $e=e_1$. This enables a simple classification of the  qualitative behavior of cosmological scenarios.

Further we impose condition $h(0)=0$ so as to avoid situations when solutions can be extended  to negative values of $e$ in the solutions of (\ref{eq5}). 
As we pointed out above, the regular solution of (\ref{eq5}) exists $\forall X\in (-\infty, \infty)$. 

Let  $e_m\ge 0$ is maximal of all roots of  $h(e)$, $h(e_m)=0$.
If, e.g.,  $h(e)>0$ for $\forall e>e_m$,  then 
any solution $e(X)$ passing through the point $X_0,e>e_m$ can be extended to all $X$-axis, it is monotonically decreasing, unbounded and it has the range  $(e_m,\infty)$.
The bounded solutions in this case are impossible. Indeed, if we suppose that $e(X)$ is bounded then, according to  Weierstrass theorem, there exists some finite value $e_\ast $, such that $e(X) \to e_\ast > e_m $ for $X \to - \infty $, whence $de/dX \to 0 $ and 
$h(e_\ast ) = 0$ contradicting to the condition that $h(e) > 0$ for $e>e_m$.

The most simple example of this case: $p=w e,\, w>-1$.

Analogously, if $h(e)<0$ for $e>e_m\ge 0$, then any solution $e(X)$ passing through the point $X_0,e>e_m$ can be extended to all $X$-axis, it is monotonically decreasing, unbounded (for large negative $X$) and it has the range  $(e_m,\infty)$.

Thus $\bf{U \downarrow }$ and $\bf{U \uparrow }$ are the only possible  types of solutions in the domain $e>e_m$.

Let $h(e_1)=0, \,h(e_2)=0$,  $e_1<e_2$. Then in the domain $e\in (e_1,e_2)$ we have either $\bf {B\uparrow }$ or  $\bf {B \downarrow }$ type depending on the sign of $h(e)$.

\section{cosmological scenarios for $k = 0, - 1$}

We are interested in the domain, range and monotonicity of the functions  $a(t),\;e(t) > 0$ satisfying equations (\ref{eq3}),(\ref{eq4}).
From  equation (\ref{eq3}) we have:
\begin{equation}
\label{eq6}
\frac{dX}{dt} = s\sqrt {\frac{8\pi }{3}e - k\exp ( - 2X)} \quad ,
\end{equation}
where $s = 1$  for the cosmological expansion ($\dot {a} > 0$) and $s = - 1$  for  the contraction ($\dot {a} < 0$).
For $k = 0, - 1$ (open or spatially-flat Universe), the right-hand side of
equation (\ref{eq3}) is always non-vanishing, and the sign $s$ does not change.

The condition, which allows us to extend solution $a(t),e(t)$ on all  values of $\left| t \right|$, is the divergence of the integral:
\begin{equation}
\label{eq7}
I(X_1 ,X_2) = \rint\limits_{X_1}^{X_2}{dX} \;\left[ {\frac{8\pi }{3}e(X) -
k\exp (-2X)} \right]^{-1/2}.
\end{equation}
both for $X_2 \to \infty $ and for $X_1 \to - \infty $, as $e(X)$ is
extended to corresponding values of the argument. If one of the conditions is
not satisfied, the solution meets a singularity and exists respectively only for $t < t_\ast < \infty $ or for $t > t_0 > - \infty $. 

For $k = - 1$, $s = 1$ all possible types of  solutions in
regions with a different signs of $h(e)$ are described in Table 1. For $k =
- 1$ the evolution always starts from finite time $t = t_0 > - \infty $, since
\begin{equation}
\label{explimit}
I(X_1 ,X_2) \le \rint\limits_{X_1}^{X_2}{dX} \;  \exp (X)= \exp (X_2)-\exp (X_1) .
\end{equation}
for all of the cases of behavior of $e(X)>0$, i.e.  the integral (\ref{eq7}) is
convergent on the lower boundary. 
Because the system involved is autonomous,  we put here and below in the latter case $t_0 = 0$.

In case of $\bf{U\downarrow} $ we get an infinite (monotonic) increasing of the
scale factor from zero to infinity and monotonic decreasing of the energy 
density $e(t) \to e_0 \ge 0$ for $t \to \infty $ (type 1.1 of Table \ref{tab1}).

If $h(e) < 0$, then we have increasing $e(t)$ and the solution either can be or cannot be extended to all times in future; in the latter case there must be a singularity of energy density at some finite value $a_0 $  of the scale factor (cf. "Big Rip"\cite{big_rip}). 
Then for $\bf{U \uparrow} $ we have two types: 1.2 --  when (\ref{eq7}) is divergent on upper boundary, $a\qty(t) \to \infty , e\qty(t) \to \infty$ as $t \to \infty$ and 1.3 -- when $X(e):-\infty \uparrow \infty$  and (\ref{eq7}) is convergent on upper limit.

For cases  $\bf{B \downarrow }$, $\bf{B \uparrow  }$ cosmological evolution continues from $t = 0$ to infinite times and energy density is always finite.
Note that if $e(t)$ tends to a finite value, then it can be identified with the current value 
of the dark energy density.

\begin{figure}[H]
    \centering
    \subfloat[Solutions for monotonic decreasing functions]{{\includegraphics[width=7cm]{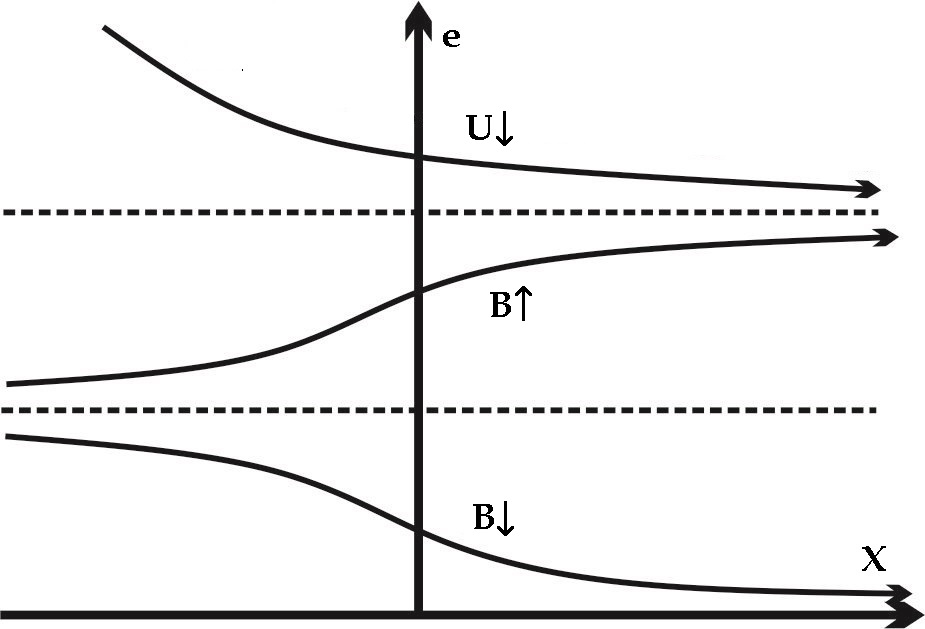} }}%
    \qquad
    \subfloat[Solutions for monotonic increasing functions]{{\includegraphics[width=7cm]{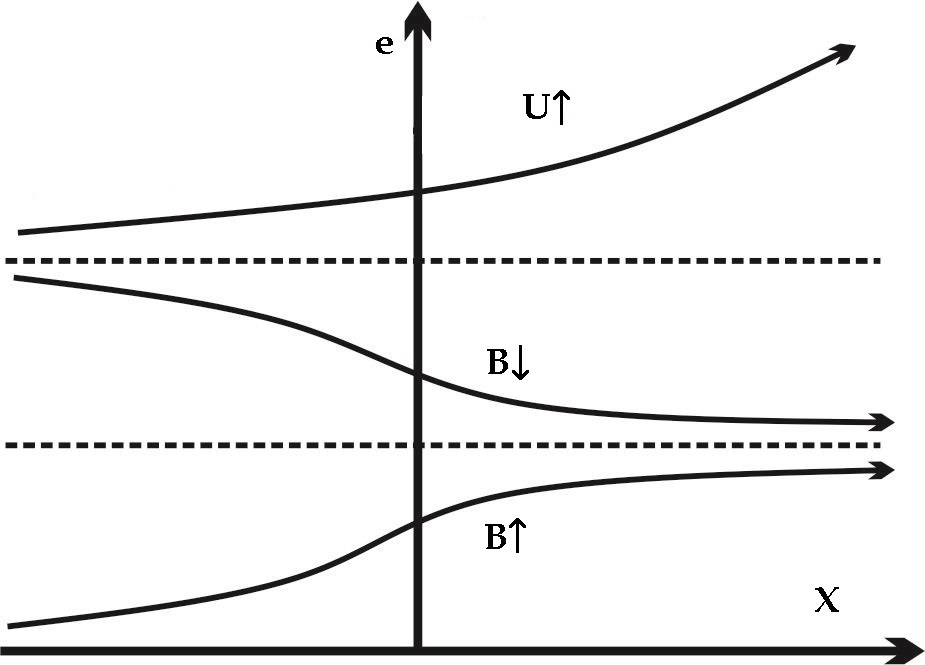} }}%
		\caption{Examples of the qualitative behavior of solutions of (\ref{eq5}) for $k=0$ and $k=-1$. The arrows show the direction of the evolution for $s=+1$.}
    \label{fig:example}%
\end{figure}

\begin{table}[tp]
		\noindent\caption{Types of qualitative behavior for $k =- 1$ ($s = 1)$ in regions with different signs $h(e)$,  $t_\ast < \infty $, $0 \le e_0 < e_1 < \infty $.}\vskip3mm\tabcolsep4.2pt \label{types_k1}
	\label{}
\begin{tabular}
{|p{28pt}|p{42pt}|p{77pt}|p{70pt}|p{70pt}|}
\hline
Type&
$e(X)$&
Domain($t)$&
$a(t)$&
$e(t)$ \\
\hline
\multicolumn{5}{|p{290pt}|}{Region $e > e_0 $: $h(e) > 0$; $h(e_0 ) = 0$}  \\
\hline
1.1&	$\bf{U \downarrow}$&	$(0,\infty )$&	$0 \uparrow \infty $&	$\infty \downarrow e_0 $ \\

\hline
\multicolumn{5}{|p{290pt}|}{Region $e > e_0 \ge 0$: $h(e) < 0$; $h(e_0 ) = 0$}  \\
\hline
1.2&	$\bf{U \uparrow}$&	$(0,\infty )$&	$0 \uparrow \infty $&	$e_0 \uparrow \infty $ \\
\hline
1.3&	$\bf{U \uparrow}$&	$(0,t_\ast )$&	$0 \uparrow \infty $&	$e_0 \uparrow \infty $ \\

\hline
\multicolumn{5}{|p{290pt}|}{Region $e \in (e_0 ,e_1 )$: $h(e) > 0$; $h(e_0 ) = h(e_1 ) = 0$}  \\
\hline
1.4&	$\bf{B \uparrow}$&	$(0,\infty )$&	$0 \uparrow \infty $&	$e_0 \uparrow e_1 $ \\
\hline
\multicolumn{5}{|p{290pt}|}{Region $e \in (e_0 ,e_1 )$: $h(e) < 0$; $h(e_0 ) = h(e_1 ) = 0$,}  \\
\hline
1.5&	$\bf{B \downarrow}$&	$(0,\infty )$&	$0 \uparrow \infty $&	$e_1 \downarrow e_0 $ \\
\hline
\end{tabular}
\label{tab1}
\end{table}

\begin{table}[!http] 
	\noindent\caption{Types of qualitative behavior for $k = 0$ ($s = 1)$ in regions with
		different signs $h(e)$, $t_\ast < \infty $, $0 \le e_0 < e_1 < \infty .
		$}\vskip3mm\tabcolsep4.2pt \label{types_k01}
\begin{tabular}
{|p{28pt}|p{42pt}|p{50pt}|p{50pt}|p{50pt}|}
\hline
Type&
$e(X)$&
Domain ($t)$ &
$a(t)$&
$e(t)$ \\
\hline
\multicolumn{5}{|p{200pt}|}{$e > e_0 :
\quad	h(e) > 0;	\quad	h(e_0 ) = 0$}  \\
\hline
2.1&	$\bf{U \downarrow }$&	$( - \infty ,\infty )$&	$0 \uparrow \infty $&	$\infty \downarrow e_0 $ \\
\hline
2.2&	$\bf{U \downarrow }$&	$(0,\infty )$&	$0 \uparrow \infty $&	$\infty \downarrow e_0 $ \\

\hline
\multicolumn{5}{|p{200pt}|}{$e > e_0 \ge 0:
\quad	h(e) < 0;	\quad	h(e_0 ) = 0$}  \\
\hline
2.3&	$\bf{U \uparrow }$&	$( - \infty ,\infty )$&	$0 \uparrow \infty $&	$e_0 \uparrow \infty $ \\
\hline
2.4&	$\bf{U \uparrow }$&	$( - \infty ,t_\ast )$&	$0 \uparrow \infty $&	$e_0 \uparrow \infty $ \\

\hline
\multicolumn{5}{|p{200pt}|}{$e \in (e_0 ,e_1 ):
\quad	h(e) > 0;	\quad	h(e_0 ) = h(e_1 ) = 0$}  \\
\hline
2.5&	$\bf{B \uparrow }$&	$( - \infty ,\infty )$&	$0 \uparrow \infty $&	$e_0 \uparrow e_1 $ \\
\hline
\multicolumn{5}{|p{200pt}|}{$e \in (e_0 ,e_1 ):
\quad	h(e) < 0;	\quad	h(e_0 ) = h(e_1 ) = 0,$}  \\
\hline
2.6&	$\bf{B \downarrow }$&	$( - \infty ,\infty )$&	$0 \uparrow \infty $&	$e_1 \downarrow e_0 $ \\
\hline
\end{tabular}
\label{tab2}
\end{table}

In case of spatially-flat Universe ($k = 0$) we do not have an estimate like (\ref{explimit}); then for certain equations of state there are 
solutions that can be extended to $t \to - \infty $, since there are cases when both $I(X_1 ,X_2)$ is divergent on the lower limit. 
All possible cases for $k = 0$ are presented in Table \ref{tab2}. The example of the case 2.1  (Table \ref{tab2}) is given by   $p(e)= -e+(e-e_0)\sqrt{e_1/(e+e_1)}, \,\,e_0\ge 0,\,e_1>0$, and the example of the case 2.2: $p= we,\,\,w>-1$.

The solutions that correspond to contracting Universe ($s = - 1)$ for $k = 0, - 1$ are obtained from the previous considerations by the change $t \to - t$.

\section{cosmological scenarios for $k = 1$}
For $k = 1$ the Universe is closed and its evolution depends on zeros of the function $F(X) = (8\pi/3) e(X)  - \exp(-2X)$. First we must separate degenerate cases when $F(X_0) = 0,\, F'(X_0 ) = 0$.
In these cases we have $F'(X) = (8\pi/3) e'(X)  +2\exp(-2X)=-(8\pi/3)(e+3p)=0$ at $X=X_0$. It is easy to see that there is  
the  solution  $a(t) \equiv \exp (X_0 )$ of the Friedmann equations (\ref{eq2},\ref{eq3}); also, there are solutions such that the point $X_0$ is an attractor or repeller: $X(t)\to X_0$ for $t\to \infty$ or/and $t\to -\infty$.

\begin{table}[bp]
		\noindent\caption{Types of qualitative behavior for $k = 1$, $F(X) > 0$. Here  $X_r$ -- finite roots of $F(X)$, $a_r = \exp X_r$.} \vskip3mm\tabcolsep4.2pt \label{types_k011}
\begin{tabular}
{|p{35pt}|p{35pt}|p{55pt}|p{55pt}|p{65pt}|}
\hline
Type №&  $e(X)$&
Domain $(t)$&
$a(t)$&
$e(t)$ \\
\hline
\multicolumn{5}{|p{300pt}|}{$\quad e(X)>e_0: h(e(X)) > 0,\,	h(e_0 ) = 0$; no zeros of $F(X) > 0$.}  \\
\hline
3.1&	$\bf{U \downarrow }$&	$( - \infty ,\infty )$&	$0 \uparrow \infty $&	$\infty \downarrow e_0 $ \\
\hline
3.2&	$\bf{U \downarrow }$&	$(0,\infty )$&	$0 \uparrow \infty $&	$\infty \downarrow e_0 $ \\

\hline
\multicolumn{5}{|p{300pt}|}{$e(X)>e(X_r)>e_0: h(e(X)) > 0,\, h(e_0 ) = 0; 	X < X_r, F(X_r ) = 0	$.}  \\
\hline
3.3&	$\bf{U \downarrow }$&	$( - \infty ,\infty )$&	$0 \uparrow a_r \downarrow 0$&	$\infty \downarrow e(X_r )	\uparrow \infty $ \\
\hline
3.4&	$\bf{U \downarrow }$&	$( - t_\ast ,t_\ast )$,\par	$t_\ast<\infty $&	$0 \uparrow a_r \downarrow 0$&
$\infty \downarrow e(X_r ) \uparrow \infty $ \\
\hline
\multicolumn{5}{|p{300pt}|}{$e_0 < e(X) < e(X_r ),
\quad	h(e(X)) > 0, h(e_0 ) = 0;\, X > X_r,	F(X_r ) = 0	$.}  \\
\hline
3.5&	$\bf{U \downarrow ,}$ \par $\bf{B \downarrow }$&	$( - \infty ,\infty )$, \par $( - \infty ,\infty )$ &	$\infty \downarrow a_r \uparrow \infty $, \par $\infty \downarrow a_r \uparrow \infty $ &	$e_0 \uparrow e(X_r ) \downarrow e_0 $, \par $e_1 \uparrow e(X_r ) \downarrow e_1$ \\
\hline
\multicolumn{5}{|p{300pt}|}{ $e_0 < e(X) < e(X_r )$, $h(e) < 0,h(e_0 ) = 0 $; $X > X_r $, $F(X_r ) = 0$.}  \\
\hline
3.6&	$\bf{U \uparrow }$&	$( - \infty ,\infty )$&	$\infty \downarrow a_r \uparrow \infty $&	$\infty \downarrow e(X_r ) \uparrow \infty $ \\
\hline
3.7&	$\bf{U \uparrow }$&	$( - t_\ast ,t_\ast )$,\par $0<t_\ast<\infty $ &	$\infty \downarrow a_r \uparrow \infty $ \par &	$\infty \downarrow e(X_r ) \uparrow \infty $ \\

\hline
\multicolumn{5}{|p{300pt}|}{$e(X_r ) < e < e_0 ,
\quad	h(e) < 0,	\quad	X > X_r ;	\quad	F(X_r ) = 0,	\quad	h(e_0 ) = 0.$}  \\
\hline
3.8&	$\bf{B \uparrow }$&	$( - \infty ,\infty )$&	$\infty \downarrow a_r \uparrow \infty $&	$e_0 \downarrow e_r \uparrow e_0 $ \\
\hline
\multicolumn{5}{|p{300pt}|}{Region $X \in \left( {X_{r_1 } ,X_{r_2 } } \right)$, $F(X_{r_1 } ) = F(X_{r_2 } ) = 0$, $h(e(X)) > 0$.}  \\
\hline
3.9&	$\bf{U \downarrow, B \downarrow }$&	$( - \infty ,\infty )$&	Oscillating&	Oscillating \\
\hline
\end{tabular}
\label{tab3}
\end{table}

\begin{figure*}[ht]
    \centering
     \subfloat[(a): types 3.1-3.2, (b): types 3.3-3.4][\centering (a): types 3.1-3.2, (b): types 3.3-3.4]{\includegraphics[width=5cm,height=3.5cm]{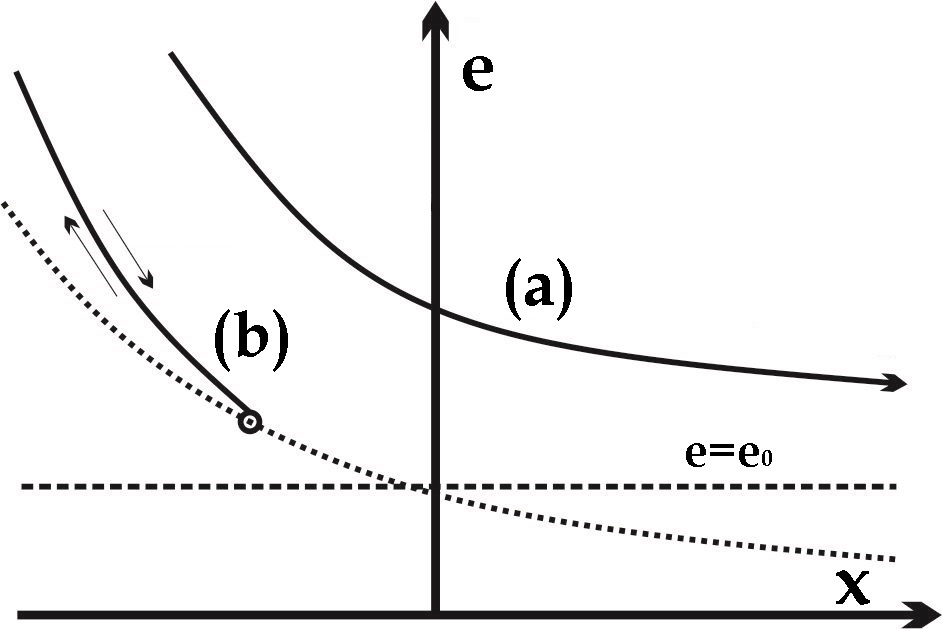} }%
     \qquad
    \subfloat[ (a): type 3.5, (b): type 3.9, (c): type 3.8][\centering  (a): type 3.5, (b): type 3.9,\par (c): type 3.8]{\includegraphics[width=5cm,height=3.5cm]{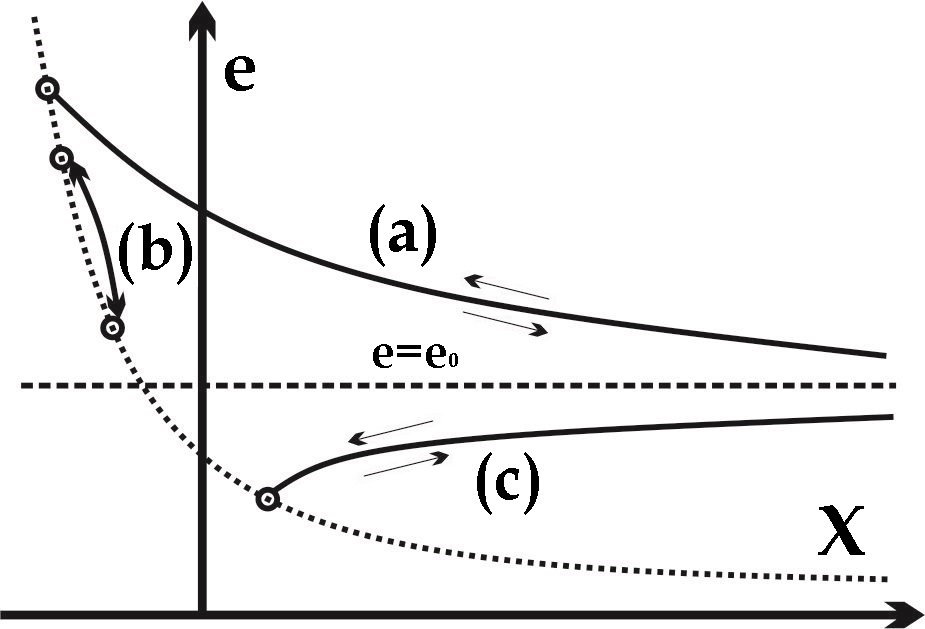} }%
    \qquad
   		\subfloat[(a): types 3.6-3.7, (b): type 3.5, (c): type 3.9][\centering  (a): types 3.6-3.7, (b): type 3.5,\par (c): type 3.9]{\includegraphics[width=5cm,height=3.5cm]{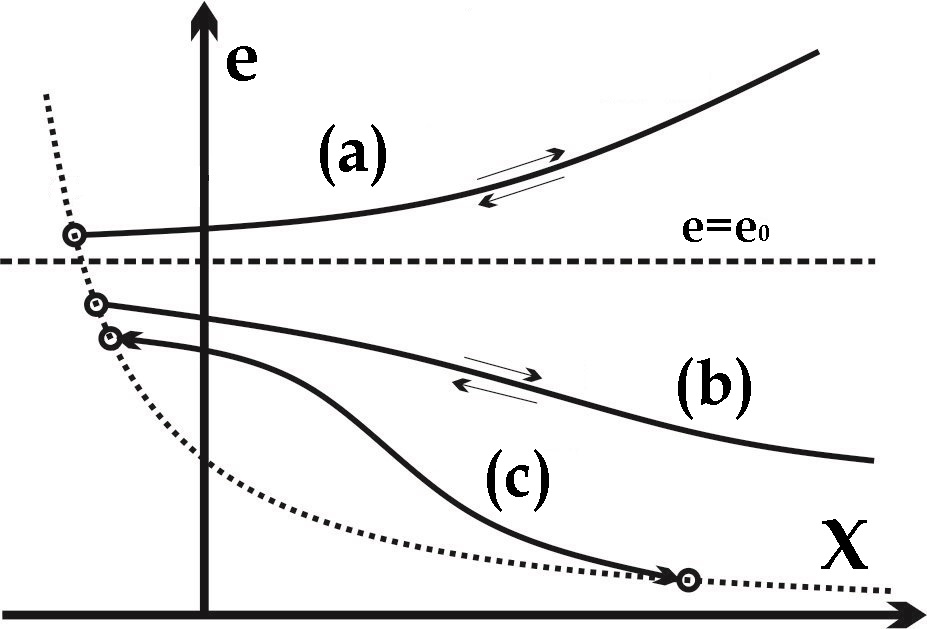} }%
		\caption{Examples of the qualitative behavior of solutions of (\ref{eq5}) for $k=1$; the arrows show the possible directions of evolution. The small rings indicate the turning points; the occurrence of two such points corresponds to periodic solutions.}
    \label{fig:example2}%
\end{figure*}

Further we confine ourselves to the case when  zeros of $F(X)$ either do not exist or they are simple: $F(X_0)=0,\, F'(X_0 ) \ne 0$; in the latter case these zeros are turning points for $X(t)$ where the change between expansion and contraction occurs (the change of sign of $s$).  
\\

Consider first the case $\bf{U \downarrow }$ of the unbounded $e(X)$. Let $h(e) > 0$ for $e>e_0$, $h(e_0 ) = 0$.

 In the region $e>e_0$, if $\forall X: F(X) > 0$  and  for  all  $X$, then only the types  analogous to 2.1, 2.2 of $k=0$ (see 3.1, 3.2 of Table \ref{tab3}) are possible.  The cases of expansion
($s = 1)$ and contraction ($s = - 1)$ are related by means of the change $t \to - t$. The cases analogous to 2.3, 2.4 are impossible here, because in the region $e>e_0$,  $h(e_0)=0$, in case of $U\uparrow$ there must be a root of  $F(X)$.

Let there is a root $X_r: F(X_r)=0$ and $F(X) > 0,\quad X < X_r < \infty$. Then the expansion is followed by contraction ($s = 1 \to s = - 1)$ at the turning point $a_r = \exp X_r$. The evolution  starts with an infinite
density and ends similarly. However, the behavior for $X\to -\infty$ is similar to previous cases 3.1,3.2: there can be either a solution that is extended for infinite times $t \to \pm \infty$ (type 3.3 of Table \ref{tab3}) or a solution with singularities at $t = \pm t_*, |t_*|<\infty$   (type 3.4).

If $F(X) > 0,\quad X > X_r > - \infty $  then we have an  evolution from  $t= - \infty $  to $t= \infty $ with a 
bounce at $X=X_r$; here we have a change from contraction to expansion and the energy density is always bounded (type 3.5 of Table \ref{tab3}).

For $\bf{U \uparrow} $ type, it is easy to see that necessarily there is  a root $X_r: F(X_r)=0$ and $F(X) > 0,\quad X > X_r >-\infty$.  We have an evolution
with the bounce with a different types of behavior from contraction to infinite expansion depending on the rate of increase of $e(X)$: a type  when solutions are defined for all $t$ and a type with the "Big Rip" in the future and in the past (see 3.6, 3.7).

In case of $\bf{B \downarrow  }$  there is always at least one root $X_r: F(X_r)=0$. Consider the domain $e\in (e_1,e_2)$, $h(e_1)=h(e_2)=0$,  $e_1<e(X_r)<e_2$.  Then we have the same qualitative behavior as in the case $\bf{U \downarrow}$, 3.5.

In case of $\bf{B \uparrow  }$  there can be only one root $X_r: F(X_r)=0$. We have an evolution $\forall t$
with the bounce from  contraction to infinite expansion; the energy density is always finite. 

At last, let $X_0 < X_1 $ be  finite roots of the function $F(X)$: $F(X_i ) =
0,\;F'(X_i ) \ne 0,\;i = 0,1$, such that $F(X) > 0$ for $X_0 < X < X_1$. This can be only either in case of $\bf{U
	\downarrow }$ or  $\bf{B \downarrow }$. Here we have an oscillating solution $a(t),e(t)$ of the Friedmann equations.  
 
Possible types of qualitative behavior are summarized in Table 3. 

\section{some generalizations}
The consideration of previous sections can be easily generalized to the case of a multi-component fluid under assumption that different DE components do not interact with each other. In this case we have the equation (\ref{eq5}) for each component $e_n,\,n=1,...,N$ separately and the reasoning  of Section \ref{solutions_of_ref_eq5} are valid for these components. Instead of equation \ref{eq6} we have
\begin{equation}
\label{eq6total}
\frac{dX}{dt} = s\sqrt {\frac{8\pi }{3}e_{tot} - k\exp ( - 2X)}\,, \quad   e_{tot}=\sum_{n=1}^{N}e_n.
\end{equation}
 The main difference from the above discussion is that $e_{tot}(X)$ can be a non-monotonous function and it is possible that $e_{tot}(X)\to \infty$ both for $X\to \infty$ and $X\to -\infty$. For example, this can be the case  
 of $e_{1}(X)\to \infty$ for  $X\to \infty$ and $e_{2}(X)\to \infty$ for  $X\to -\infty$. In this case for $k=0$ and $k=1$  it is allowed that $a(t):0\uparrow \infty$ on a finite interval $(0,t_*), \,t_*<\infty$. This behavior has been prohibited  in case of one component. Note that here $a(t)$ is a monotonous function, in contrast to type 3.4 or 3.7 of Table \ref{tab3}.  For the one-component case with $k=-1$ such domain and range of $a(t)$ is also possible (cf. type 1.3 of Table \ref{tab1}) but here the energy density is a monotonous function. For $k=1$, the multicomponent case yields one more possible scenario with  $a(t):0\uparrow \infty$ on  $(-\infty,t_*), \,t_*<\infty$ in addition to the types of Table \ref{tab3}.
 
 The other generalization concerns the smoothness of EoS and $h(e)$. Evidently, this  requirement and the inequality (\ref{dpde}) can be replaced by a single requirement of Lipschitz continuity $|p(e_1)-p(e_2)|<K|e_1-e_2|$ for all $e\in (\-\infty,\infty)$, where $0<K<\infty$. 
  
  If this condition is violated then solutions of (\ref{eq5}) can appear with either crossing of the phantom line $e\equiv e_0$ with $h(e_0)=0$ or, e.g., with $e(X)<e_0$ for $X<X_0$ and $e(X)\equiv e_0$ for $X\ge X_0$. The simple example of the latter EoS is $p(e)=-e+C_1 \sqrt{|e-e_0|},\, C_1>0,$ with the solutions $e=e_0+(3C_1/2)^2(X-X_0)|X-X_0|$ and $e\equiv e_0$. In case of singularities of $p(e)$ even more complicated situations are possible, e.g., when the solution $e(X)$ cannot be extended for all $X$. 
  
\section{discussion}

Thus, in the framework of the hydrodynamic model of homogeneous isotropic universe
with a general smooth barotropic equation of state, we presented a classification of qualitative scenarios of cosmological
evolution. We listed all possible types of the solutions depending on whether their domains and ranges can be  finite or infinite. The classification includes the
"traditional" scenario, which starts from $t = 0 $ and continues for infinite times. Also, there is a situation
is when the time, where a solution exists, could be
limited.  However, there are equations of state that generate scenarios of the eternal Universe, which exists from the infinite times in the past, as well as scenarios for which the energy density $e(t)$ is always finite. This class includes scenarios for a closed universe with a bounce and oscillating solutions.
Note that most of the examples discussed above can be found
in the other works in the context of specific problems, in particular possible
types of singular behavior have been analyzed in \cite{Noiri_1,worse_big_rip,grand_rip,parnovsky}. However, our
classification covers all possible qualitative types of cosmological evolution from a unified viewpoint.

It is essential that, within our discussion, it is impossible to pass through any zero point of the enthalpy, e.g., from the region of 
a regular behavior where $h(e) > 0$, to the region where $h(e) < 0$, where it
is possible for singularities of the scale factor to occur in the  future. This is due to the
uniqueness of the solution of equation (\ref{eq5}) in case of smoothness of the
equation of state (and specific enthalpy respectively). Our classification does not include non-smooth equations of state, which can lead to solutions intersecting points of zero enthalpy (e.g. \cite{parnovsky}).  Consideration of such EoS may be of interest  because the
smoothness condition can be violated in phase transitions. Our qualitative analysis can be  generalized to include such
options, though when considering the global cosmological behavior, it would generate too large 
number of additional types. In the case of smooth equations of
state our classification is complete.

\section*{\sc acknowledgement}
This work has been partially supported by the State Fund of Ukraine for the Fundamental Research.


\begin{thebibliography}{3}
	{\small
		\bibitem{grt_1}~Yatskiv~Ya.S., Alexandrov~A.N., Vavilova~I.B.,   et al. General Relativity theory: tests through time. Kyiv, Akademperiodyka,  2005.
		\bibitem{grt-3}~
		Alexandrov~A.N., Vavilova~I.B., Zhdanov~V.I.,  et al. General Relativity Theory: 
		Recognition through Time.  Kiev, Naukova dumka, 2015.
		\bibitem{dedm} Novosyadlyi~B., Pelykh~V., Shtanov~Yu., Zhuk~A. Dark energy and
		dark matter of the universe: in three volumes / Ed. V.~Shulga. -- Vol. 1:
		Dark matter: Observational evidence and theoretical models. K.:
		Akademperiodyka, -- 2013.
		\bibitem{Noiri_1}Nojiri~S.,  Odintsov~S.D., Tsujikawa~S. Properties of
		singularities in (phantom) dark energy universe. Phys. Rev. ~D71, 063004 (2005).
		\bibitem{Noiri_2}Nojiri~S.,  Odintsov~S.D. Modified gravity with negative and
		positive powers of the curvature: unification of the inflation and of the
		cosmic acceleration. Phys.Rev. ~D68, 123512 (2003).
		\bibitem{worse_big_rip} Bouhmadi-L\'{o}pez~M., Gonz\'{a}lez-D\'{\i}az~P.F.,
		Mart\'{\i}n-Moruno~P. Worse than a big rip? Phys. Lett. ~B659, 1 (2008).
		\bibitem{bamba} Bamba~K., Odintsov~S.D. Universe acceleration in modified
		gravities: $F(R)$ and $F(T)$ cases [arXiv:1402.5114, hep-th] (2014).
		\bibitem{jenk_zh} Jenkovszky~L.L., Zhdanov~V.I., Stukalo~E.J. Cosmological model
		with variable vacuum pressure. Phys.Rev. ~D90, 023529 (2014).
		\bibitem{grand_rip}  Fern\'{a}ndez-Jambrina~L. Grand Rip and Grand Bang/Crunch
		cosmological singularities. Phys. Rev. ~D90, 064014 (2014).
		\bibitem{parnovsky} Parnovsky S.L. Big Rip and other singularities in isotropic
		homogeneous cosmological models with arbitrary equation of state. Odessa
		Astronomical Publications.  28/2, 137 (2015)
		\bibitem{big_rip} Caldwell~R.R., Kamionkowski~M., Weinberg~N.N. Phantom Energy:
		Dark Energy with w<-1 Causes a Cosmic Doomsday. Phys.Rev.Lett., 91, 
		071301 (2003)
	}
\end{thebibliography}
\end{document}